\documentclass[twocolumn,aps,pre]{revtex4}

\usepackage{graphicx}
\usepackage{dcolumn}
\usepackage{amsmath}
\usepackage{latexsym}
\usepackage{verbatim}
\usepackage{color}

\begin {document}

\title {$Q$-factor: A measure of competition between the topper and the average in percolation and in SOC}

\author{Asim Ghosh$^1$, S. S. Manna$^{2,}$\footnote[1]{asimghosh066@gmail.com, subhrangshu.manna@gmail.com (corresponding author), bikask.chakrabarti@saha.ac.in} and Bikas K. Chakrabarti$^{3,4}$}
\affiliation{$^1$Department of Physics, Raghunathpur College, Raghunathpur 723133, India \\
             $^2$B-1/16 East Enclave Housing, 02 Biswa Bangla Sarani, New Town, Kolkata 700163, India \\
             $^3$Saha Institute of Nuclear Physics, Kolkata 700064, India \\
             $^4$Economic Research Unit, Indian Statistical Institute, Kolkata 700108, India}

\begin{abstract}
      We define the $Q$-factor in the percolation problem as the quotient of the size of the 
   largest cluster and the average size of all clusters. As the occupation probability $p$
   is increased, the $Q$-factor for the system size $L$ grows systematically to its maximum 
   value $Q_{max}(L)$ at a specific value $p_{max}(L)$ and then gradually decays. Our numerical 
   study of site percolation problems on the square, triangular and the simple cubic lattices 
   exhibits that the asymptotic values of $p_{max}$ though close, are distinctly different 
   from the corresponding percolation thresholds of these lattices. We have also shown using 
   the scaling analysis that at $p_{max}$ the value of $Q_{max}(L)$ diverges as $L^d$ ($d$ 
   denoting the dimension of the lattice) as the system size approaches to their asymptotic 
   limit. We have further extended this idea to the non-equilibrium systems such as the 
   sandpile model of self-organized criticality. Here, the $Q(\rho,L)$-factor is the quotient 
   of the size of the largest avalanche and the cumulative average of the sizes of all the 
   avalanches; $\rho$ being the drop density of the driving mechanism. This study has been 
   prompted by some observations in Sociophysics.
\end{abstract}

\maketitle

\section{Introduction}

      Critical fluctuations of all length scales appearing at the critical points are the signatures of
   phase transitions. Over the last century, extensive studies of phase transitions have helped establishing 
   the statistical physics descriptions of the scaling theory and the critical phenomena in different
   physical systems. For example, few well studied systems are magnetic and fluid systems \cite {Stanley}, 
   polymer systems \cite {Gennes}, \cite {Stauffer} for percolating systems \cite {Stauffer}, and Self-organized
   Critical (SOC) systems \cite {Bak}. Essentially, the order parameter 
   of the corresponding systems vanish following in general a singular power law or critical behavior at the 
   critical point and beyond. Its higher moments include susceptibilities, diverge again with singular 
   or critical power law exponent values at the respective critical points. For SOC systems, these singular 
   behaviors are seen from the pre-critical side and then remains critical in the SOC state of the systems.
   For practical purposes, these diverging susceptibilities help locating the critical point.

      For social systems, scientists had studied for ages, starting with Pareto law 80-20 law \cite {Pareto}, Lorenz 
   function \cite {Lorenz}, Gini index \cite {Gini}, Hirsch index \cite {Hirsch}, etc., the extreme unequal distributions of income or wealth, 
   votes, paper citations respectively. Following some recent observations \cite {Ghosh1,Ghosh2,Ghosh3} of extreme inequality level 
   in citation statistics of successful individuals and even institutions / universities / journals, with Gini 
   and other inequality index values going beyond the Pareto 80-20 limit, we studied and found \cite {Manna-Biswas} clear 
   presence of similar level of the inequality index values in the physical models of SOC system, like the 
   Bak, Tang, Wiesenfeld (BTW) sandpile \cite {btw} and the Manna \cite {Manna} sandpile. Particularly, in our recent study 
   \cite {Ghosh4} of the citation statistics of some very successful prize winning scientists and a few others not so 
   successful scientists it has been observed that their research dynamics is clearly SOC like and most successful have 
   achieved the critical level in their citation inequalities, while others are still approaching that level, 
   though they have not reached there. All these studies showed, just the average high level of citations per 
   paper (reflected by the Hirsch index values, which are determined by the effective network coordination 
   or Dunbar number \cite {Dunbar,DunbarWiki} do not reflect the success of the scientist, but the high level (beyond the Pareto 
   level) of critical fluctuations in citations from publication to publication of the scientist. Indeed, it 
   was seen in \cite {Ghosh4} that crossing a threshold value of a simple quotient of the citation number of the highest 
   cited paper and the average citation of all the papers (including the highest cited one) by the scientist 
   gives very good correlation with the appreciations by the respective communities.

      Following this clue, we study here how the topper competes with the average in the well known models of
   percolation processes and in the sandpile model of self-organized criticality. In the percolation model we
   have defined the $Q$-factor as the quotient of the largest cluster size and the average size over all 
   clusters for the percolation models. Similarly, the $Q$-factor in the sandpile model has been defined
   as the quotient of the largest avalanche size and the average size over all avalanches. As the control
   variables are tuned in these problems $Q$-factors grow very sharply right before a specific value of 
   the control variable, reaches the maxima, and then decay very rapidly. The locations of the maxima are
   distinctly different from the critical points of these systems.

      We have described our calculations and results of the percolation problem for the square, triangular and
   the simple cubic lattices in the three sub-sections of section II. Here, we have calculated the $Q$-factors
   for the entire range of the occupation probability $p$. A nice finite-size extrapolation gives the precise 
   value of the percolation occupation probability $p_{max}$ in the asymptotic limit which we found to be larger 
   than their percolation thresholds. In section III we have executed similar analysis for the BTW sandpile where
   we used the drop density $\rho$ as the tuning parameter. The value of $\rho_{max}$ in the asymptotic limit 
   have been calculated. Finally, we have summarized in section IV.

\begin{figure}[t]
\includegraphics[width=6.0cm]{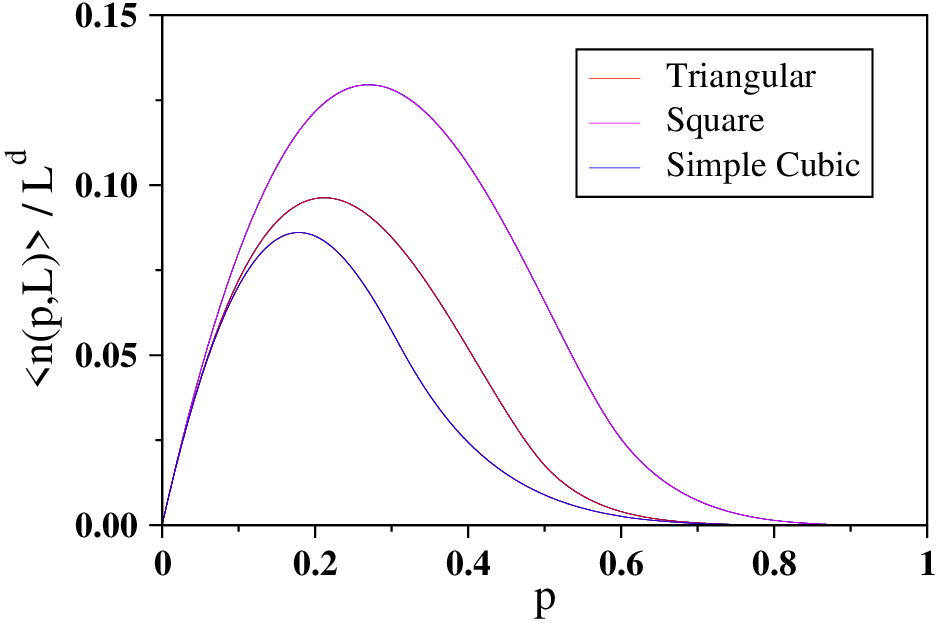}
\caption{
    Plot of the average number of distinct clusters per lattice site $\langle n(p,L) \rangle / L^d$
    against the site percolation occupation probability $p$. For each type of lattice the data for
    three different system sizes are plotted which overlapped completely, only the colors used for 
    the largest lattices are visible. 
}
\label{FIG01}
\end{figure}

\section{Site Percolation}

\subsection {Square lattice}

      An initially empty square lattice of size $L \times L$ has been gradually filled in by occupying the 
   randomly selected lattice sites one by one. At any arbitrary intermediate stage the fraction `$p$' of 
   occupied sites is referred as the percolation occupation probability. A cluster is defined as the set of 
   occupied sites connected by nearest neighbour distances. Different distinct clusters have been identified 
   using the well known Hoshen Kopelman algorithm \cite {Hoshen}. Since we are not going to study any spanning property
   of the percolation clusters we have used the periodic boundary conditions in all simulations reported here. 
   The number of distinct clusters $n(p,L)$ increases from unity at $p \to 0+$, reaches a maximum at some 
   intermediate $p$ value and then finally goes down to unity again at $p = 1$. We refer the entire process as
   a `run'. 

      In Fig. \ref {FIG01} we have plotted the average number of distinct clusters per lattice site $\langle n(p,L) 
   \rangle / L^d$ against $p$ for three different system sizes of the square, triangular, and the simple cubic 
   lattices, where $d$ represents the Euclidean dimensional of these lattices. The collapse of the plots on top
   of one another for three system sizes is extremely well. The sizes of the lattices used are $L$ = 256, 1024, and 
   4096 for the square and triangular lattices, where as $L$ = 32, 64, and 128 for the simple cubic lattice. The 
   peak positions of these curves have coordinates: (0.26968, 0.12954) for the square lattice, (0.21192, 0.096306) 
   for the triangular lattice, and (0.17871, 0.086066) for the simple cubic lattice.

\begin{figure}[t]
\includegraphics[width=6.0cm]{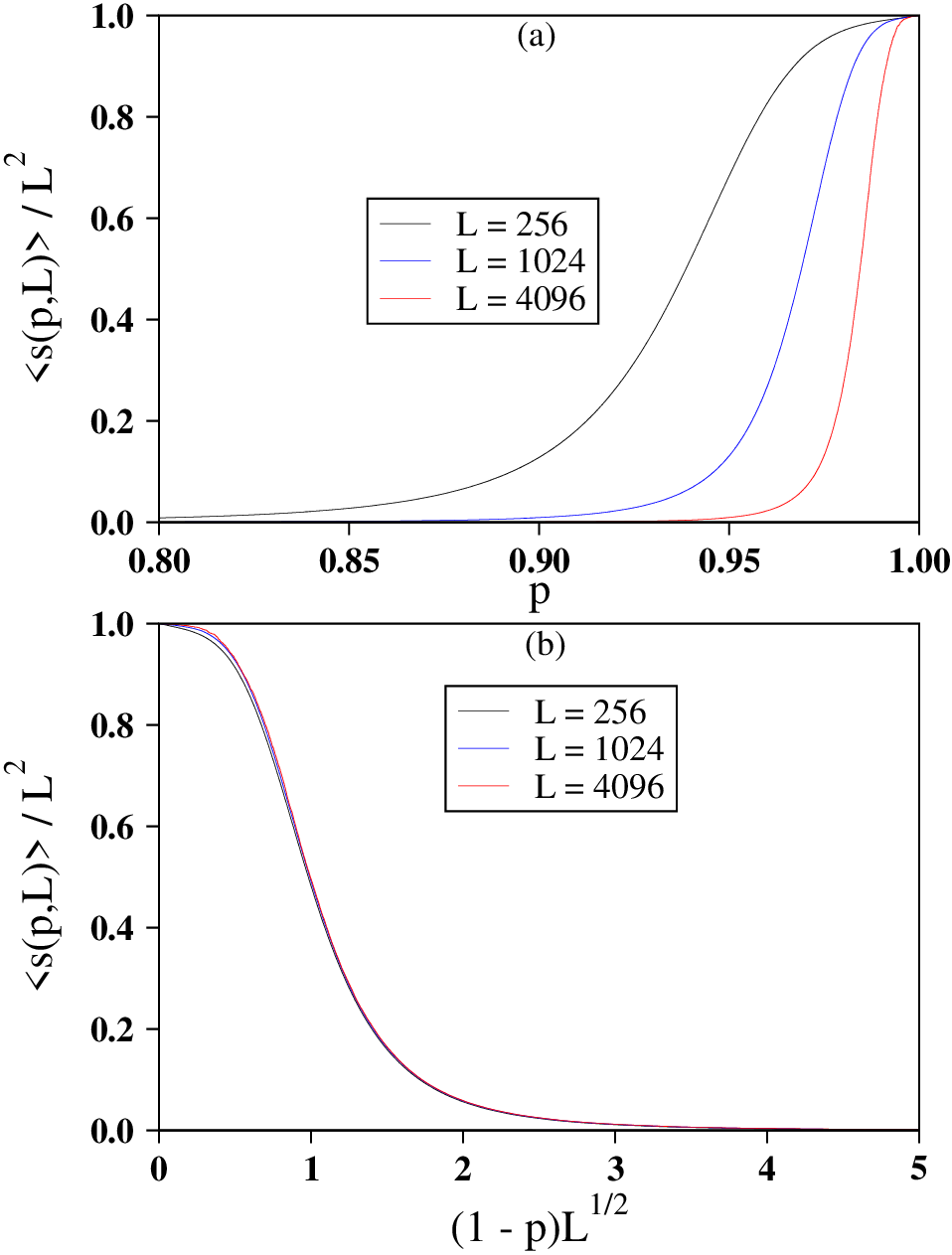}
\caption{
   (a) Plot of the average cluster size $\langle s(p,L) \rangle$ scaled by the total number $L^2$ of 
   lattice sites against the site percolation occupation probability $p$ for the square lattice.
   (b) The same data have been replotted against $(1 - p)L^{1/2}$ which yield nice collapse of the
   data.
}
\label{FIG02}
\end{figure}

      At an intermediate stage the average size of all clusters including the largest one is therefore 
   $s_{av}(p,L) = pL^d / n(p,L)$. This is further averaged over a large number of independent runs and we define 
   the average cluster size $\langle s(p,L) \rangle = \langle s_{av}(p,L) \rangle$. In Fig. \ref {FIG02}(a) we have 
   plotted the scaled average cluster size $\langle s(p,L) \rangle / L^2$ against $p$ only for the square lattice and 
   again for the same three system sizes. The curves become sharper as $p \to 1$ and as the system size becomes larger. 
   In Fig. \ref {FIG02}(b) we have inverted the $x$-axis and re-plotted the same data $\langle s(p,L) \rangle / L^2$ 
   against $(1 - p)L^{1/2}$ to observe a nice collapse of the data over the entire range of $p$ values.

\begin{figure}[t]
\includegraphics[width=6.0cm]{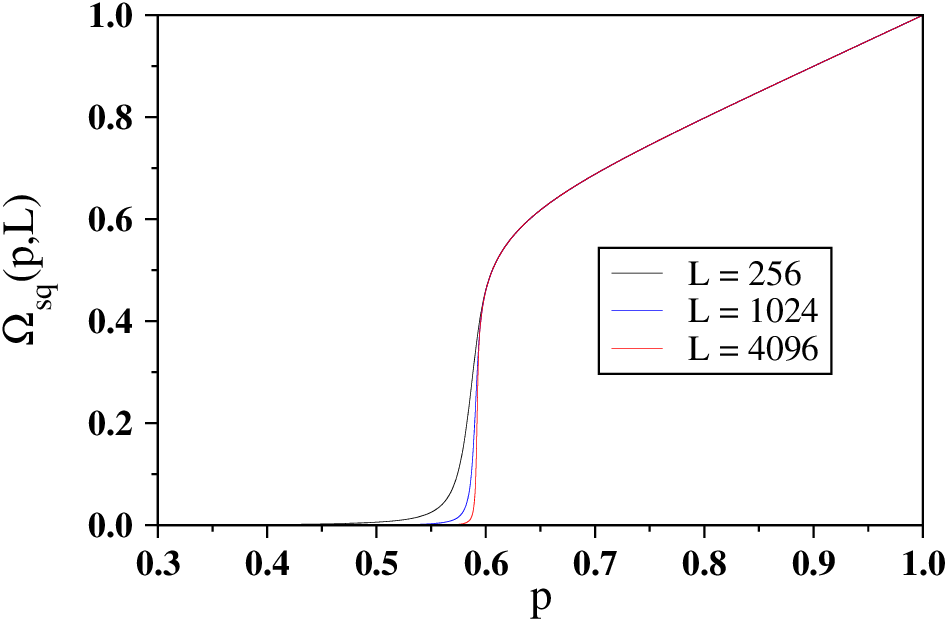}
\caption{
   Plot of the percolation order parameter $\Omega_{sq}(p,L) = \langle s_{max}(p,L) \rangle / L^2$ against the site
percolation occupation probability $p$ for the square lattice.
}
\label{FIG03}
\end{figure}

      As more and more sites are occupied, the growth of the size $s_{max}$ of the largest cluster has 
   been monitored. The order parameter $\Omega(p,L)$ of the percolation transition is defined as the 
   fractional size of the largest cluster averaged over many independent runs, i.e., $\Omega(p,L) = \langle s_{max}(p,L) \rangle / L^d$.
   In Fig. \ref {FIG03} we have plotted the order parameter $\Omega_{sq}(p,L)$ for the square lattice
   for three different system sizes. Larger the system size, the growth of the order parameter becomes 
   sharper. For a particular system $\Omega_{sq}(p,L)$ grows rapidly as the percolation occupation 
   probability $p$ approaches from below the site percolation threshold of the square lattice whose best 
   value till date is $p_c(sq) = 0.59274605079210(2)$ \cite {Jacobsen,Ziff-Wiki}.

\begin{figure}[t]
\includegraphics[width=6.0cm]{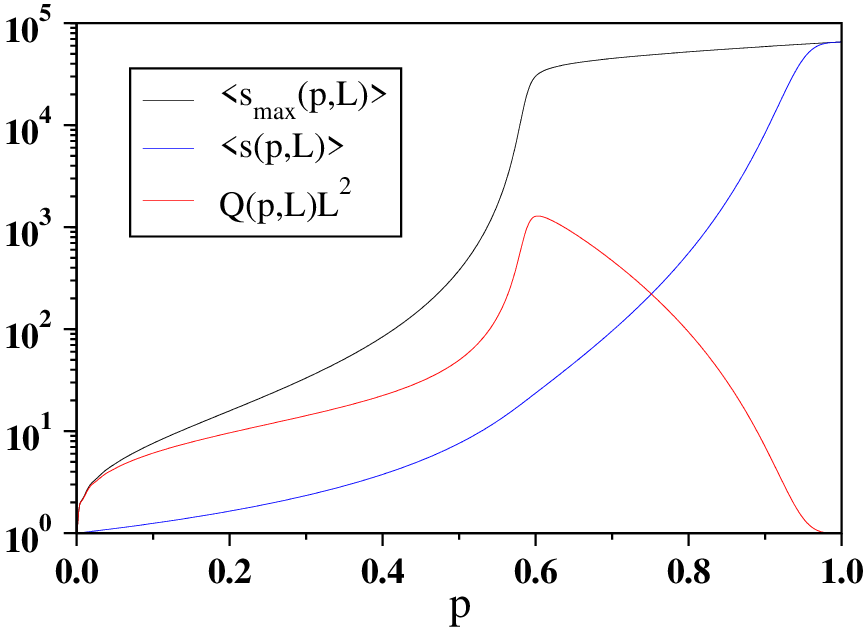}
\caption{
      Plot of the average size of the largest cluster $\langle s_{max}(p,L) \rangle$,
   average size of all clusters $\langle s(p,L) \rangle$, and the $Q$-factor $Q(p,L)L^2$
   against the site occupation probability $p$ for a system of size $L$ = 256 on the 
   square lattice.
}
\label{FIG04}
\end{figure}
\begin{figure}[t]
\includegraphics[width=6.0cm]{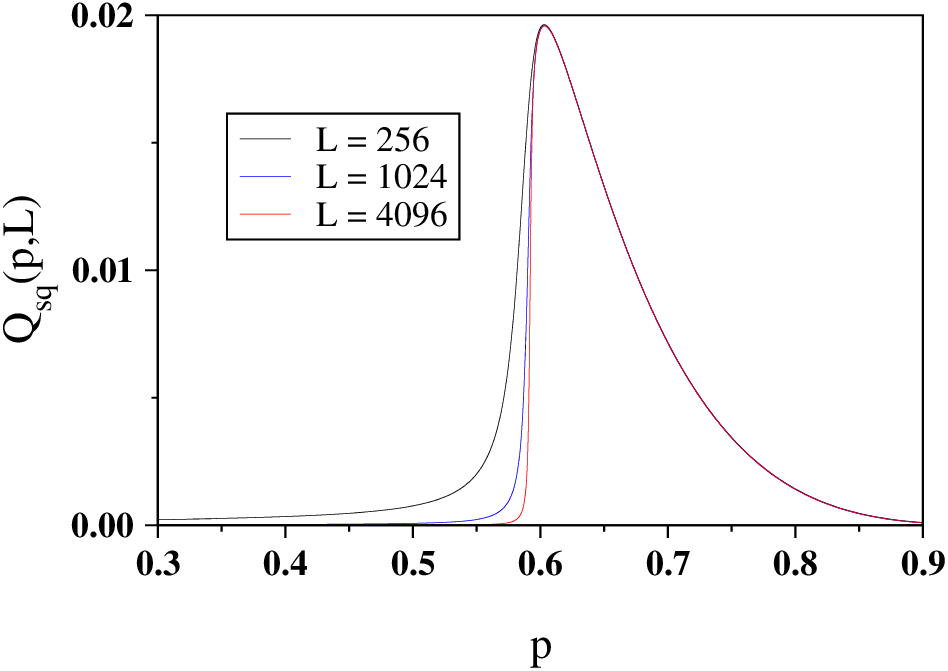}
\caption{
   Plot of $Q_{sq}(p,L)$ against the percolation occupation probability $p$ for the square lattice.
}
\label{FIG05}
\end{figure}

      Now we define the $Q$-factor as the quotient of the average size $\langle s_{max}(p,L) \rangle$ of the
   largest cluster and the average size $\langle s(p,L) \rangle$ of all clusters for every value of $p$ for a 
   certain system size $L$ as:
\begin{equation}
	\begin{tabular}{lcl}
		$Q(p,L)$ &=& $(\langle s_{max}(p,L) \rangle / \langle s(p,L) \rangle) / L^d$ \\
		         &=& $\Omega(p,L) / \langle s(p,L) \rangle$.
        \end{tabular}
\label {EQN01}
\end{equation}
   We have plotted in Fig. \ref {FIG04} three quantities for a particular system size $L$ = 256.
   They are, (i) the average size $\langle s_{max}(p,L) \rangle$ of the largest cluster, (ii) the average size
   $\langle s(p,L) \rangle$ of all clusters, and (iii) the $Q(p,L)$ factor multiplied by the system size $L^2$.
   The first two quantities are monotonically increasing functions of $p$. It is observed that when $p$ gradually increases to 
   a specific value $\langle p_{max}(L) \rangle$, the value of $\langle s_{max}(p,L) \rangle$ becomes
   increasingly larger than the average cluster size $\langle s(p,L) \rangle$ and therefore $Q(p,L)$
   increases very sharply. However, after crossing $\langle p_{max}(L) \rangle$, the growth of $\langle s_{max}(p,L) \rangle$
   becomes slower but $\langle s(p,L) \rangle$ maintains its previous growth rate. Consequently, their ratio $Q$-factor decays
   gradually which explains the existence of a peak of $Q$ at $\langle p_{max}(L) \rangle$. This is visible in Fig. \ref {FIG05} 
   where we have plotted $Q_{sq}(p,L)$ 
   against $p$ for three different system sizes. All three curves have single peaks of nearly the same heights 
   but their positions have systematic variations. 

\begin{figure}[t]
\includegraphics[width=6.0cm]{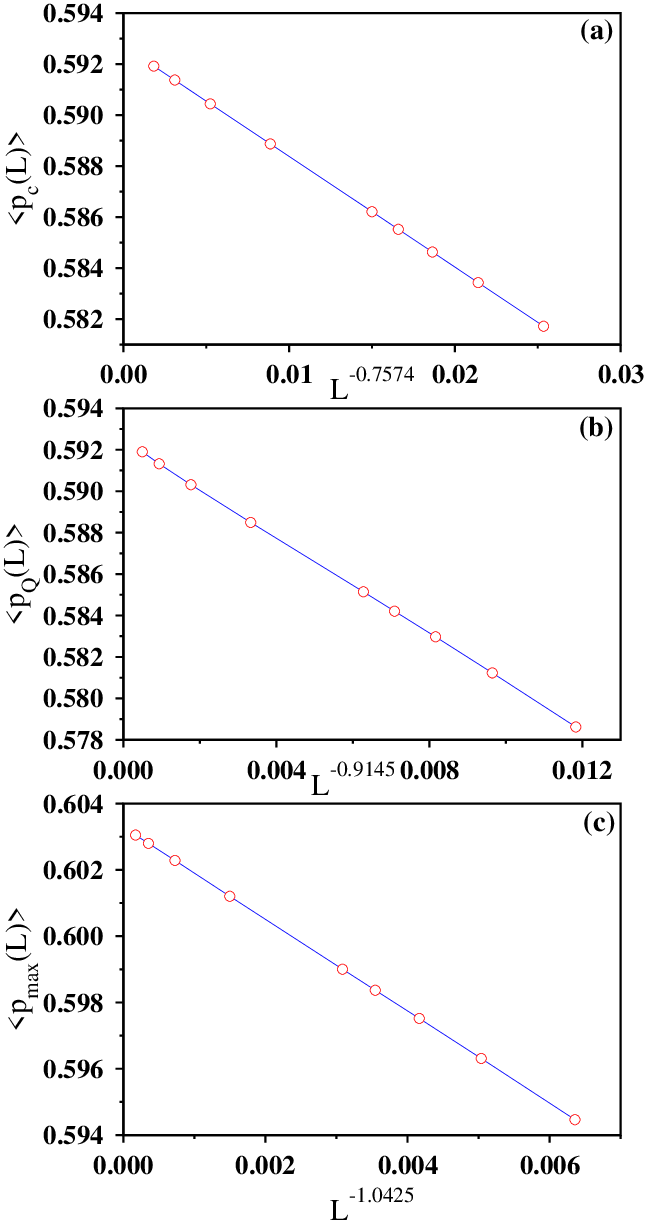}
\caption{Finite size extrapolations with suitable tuning parameters yield
(a) $p_c$ = 0.592717, $1/\nu_1$ = 0.7574; (b) $p_Q$ = 0.592419, $1/\nu_2$ = 0.9145 and (c) $p_{max}$ = 0.603288, $1/\nu_3$ = 1.0425.
}
\label{FIG06}
\end{figure}

      Now we present numerical evidences in Fig. \ref {FIG06} to claim that the asymptotic value 
   $p_{max} = \lim_{L\to\infty} \langle p_{max}(L) \rangle$
   is distinctly different from the ordinary percolation threshold $p_c$ on the same lattice. Let us denote a typical run from 
   an empty lattice ($p$ = 0) to a fully occupied lattice ($p$ = 1) on the square lattice of size $L \times L$ by $\alpha$. For every 
   $\alpha$ we estimate three different values of the occupation probability, namely:
   (i) the value of occupation probability $p_c(\alpha,L)$ at which the occupation of only the next site in the sequence causes
   the maximal jump of the size of the largest cluster $s_{max}(\alpha,L)$;
   (ii) the value of $p_Q(\alpha,L)$ at which the occupation of only the next site in the sequence causes the maximal jump of
   the value of the $Q(\alpha,L)$-factor, and
   (iii) the value of $p_{max}(\alpha,L)$ at which the ratio $s_{max}(\alpha,p,L) / s_{av}(\alpha,p,L)$ reaches its maximum value.

      Their average values $\langle p_c(L) \rangle$, $\langle p_Q(L) \rangle$ and $\langle p_{max}(L) \rangle$ have been calculated 
   over a large number of runs, namely $10^8$ runs for lattices of size up to $L$ = 128, which decreases to 18000 for $L$ = 4096. 
   Each of these quantities is then extrapolated using a finite size correction term in the power law form: 
   $\langle p_c(L) \rangle = p_c - AL^{-1/\nu_k}$ 
   where $\nu_1$, $\nu_2$ and $\nu_3$ correspond to $p_c$, $p_Q$ and $p_{max}$ respectively. Here, $\nu_1=\nu$ the ordinary correlation 
   length exponent of two dimensional percolation problem. In comparison to $1/\nu_1=0.75$ we get 0.7574, which is quite close. The values 
   of $1/\nu_2 = 0.9145$ and $1/\nu_3 = 1.0425$ show that the values of $\nu_2$ and $\nu_3$ are quite different from $\nu$, but 
   their values are close to each other and nearly equal to 1.

      After extrapolation the asymptotic value of $p_c = 0.592717$ has been obtained which is very close to the actual value of the
   site percolation threshold $\approx$ 0.592746 with a difference of $\approx 0.00003$. The asymptotic value of $p_Q=0.592419$ has been
   found which is $\approx 0.0003$ away from the percolation threshold. Where as, the asymptotic value of $p_{max} = 0.603288$
   differs by an amount $\approx 0.01$ from the percolation threshold. With this analysis, we conclude that while the values of
   $p_Q$ and $p_c$ are most likely to be the same, the value of $p_{max}$ is infact, distinctly different from $p_c$.

      In our next analysis we have calculated and plotted the ratios of successive values of three quantities. Specifically,
   if the occupation probability is increased by a small amount of say $\Delta p = 1/L^2$, i.e., one more site is occupied,
   then to what factors the quantities (i) $\langle s_{max}(p,L) \rangle$, (ii) $\langle s(p,L) \rangle$ and (iii)
   $Q(p,L)$ are increased? Let these ratios be denoted by

   {\centerline {${\cal R}[\langle s_{max}(p,L) \rangle] = \frac{\langle s_{max}(p+\Delta p,L) \rangle}{\langle s_{max}(p,L) \rangle}$,}}

   {\centerline {${\cal R}[\langle s(p,L)  \rangle] = \frac{\langle s(p+\Delta p,L) \rangle}{\langle s(p,L) \rangle}$,}}

\begin{figure}[t]
\includegraphics[width=7.5cm]{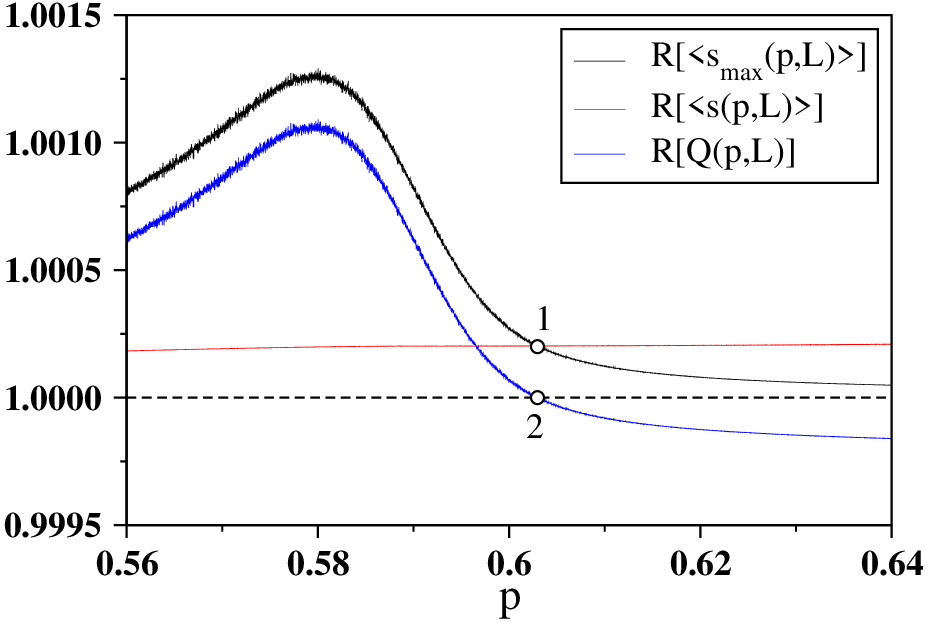}
\caption{
      Plots of ${\cal R}[\langle s_{max}(p,L) \rangle]$, ${\cal R}[\langle s(p,L) \rangle]$ and ${\cal R}[Q(p,L)]$
   against the site occupation probability $p$ for the square lattice of size $L$ = 256. The first two curves 
   meet at the point 1 where their values are equal. Therefore, their ratio is unity which corresponds to 
   ${\cal R}[Q(p,L)]=1$ at the point 2.
}
\label{FIG07}
\end{figure}

   {\centerline {${\cal R}[Q(p,L)] = \frac{\langle Q(p+\Delta p,L) \rangle}{\langle Q(p,L) \rangle} =
   \frac{{\cal R}[\langle s_{max}(p,L) \rangle]}{{\cal R}[\langle s(p,L)  \rangle]}$}}
   respectively and are plotted in Fig. \ref {FIG07} for $L$ = 256. In Fig. \ref {FIG03} we find the size of the largest cluster 
   increases very fast right before the percolation threshold, but right after the percolation it starts increasing with $p$ 
   approximately linearly. Therefore the ${\cal R}[\langle s_{max}(p,L) \rangle]$ must be having a peak at $p_c(L)$ and the black 
   curve indeed shows a peak at $p_c(L=256) \approx 0.579815$. From Fig. \ref {FIG02} we observed that the value of
   $\langle s(p,L) \rangle$ increases very slowly except when $p$ is nearly equal to 1. Therefore, the curve in red in Fig. 
   \ref {FIG07} exhibits the slow variation of ${\cal R}[\langle s(p,L) \rangle]$ against $p$. Consequently, their ratio 
   ${\cal R}[Q(p,L)]$ (in blue) is also having a peak at $p_c$, and beyond this peak, it decreases systematically. 

      Two points are to be noticed: The $Q$-factor has
   its maximum at $p_{max}(L)$ where the ratio ${\cal R}[Q(p,L)]$ is equal to unity. Therefore, at $p=p_{max}(L)$ the two curves
   meet at point 1 where ${\cal R}[\langle s_{max}(p,L) \rangle]$ = ${\cal R}[\langle s_{av}(p,L) \rangle]$.
   The other point 2 on the ${\cal R}[Q(p,L)]$ against $p$ plot represents the point ${\cal R}[Q(p_{max},L)=1]$. We argue, if $p_c(L)$
   and $p_{max}(L)$ both assume the same asymptotic value i.e., $p_c=p_{max}$, it would mean a discontinuous drop in the
   values of ${\cal R}[Q(p)]$ at this value of $p$, which is not possible since both the largest and the average cluster sizes vary continuously
   in a continuous phase transition like the ordinary percolation.

\begin{figure}[t]
\includegraphics[width=6.0cm]{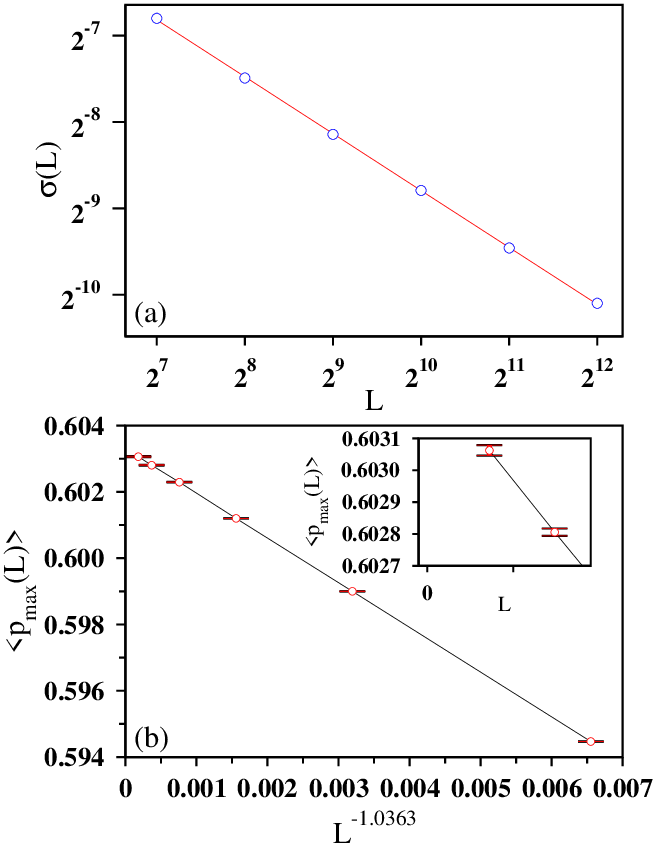} 
\caption{
(a) The standard deviation $\sigma(L)$ for the values of $p_{max}(L)$ of the square lattice have been plotted
    against the system size $L$ on the $\log - \log$ scale. The estimation of slope implies $\sigma(L) \sim L^{-0.658}$.
(b) The average values of $\langle p_{max}(L) \rangle$ have been plotted against $L^{-1.0363}$ to obtain a nice straight line.
    Each point is marked with its error bar. The extrapolated value of $p_{max} = 0.6033 \pm 0.0002$ has been obtained.
}
\label{FIG08}
\end{figure}

      We have calculated the error in our estimate for the asymptotic value of $p_{max}$. For a system of size $L$
   we have calculated the standard deviation $\sigma(L) = \{\langle p^2_{max}(L) \rangle - \langle p_{max}(L) \rangle^2\}^{1/2}$.
   We have plotted in Fig. \ref {FIG08}(a) the values of $\sigma(L)$ against $L$ using the double-logarithmic scale. We have observed that
   $\sigma(L)$ nicely scales as $L^{-0.658}$. Let us denote the number of independent runs be $M$, then we define
   the error as $e(L) = \sigma(L) / M^{1/2}$. For this plot the number of runs $M$ varied from 24 million for $L$ = 128 to 3000 for $L$ = 4096.
   In the Fig. \ref {FIG08}(b) of $\langle p_{max}(L) \rangle$ values have been plotted against $L^{-1.0363}$ and we have plotted 
   errors using the vertical lines. For each point we have drawn a vertical line from $\langle p_{max}(L) \rangle - e(L)$ to 
   $\langle p_{max}(L) \rangle + e(L)$ and then two
   horizontal bars of fixed length at the two ends of the vertical line (a zoomed plot of only the two points for $L$ = 2048 and 4096 has
   been shown for clarity in the inset). It is obvious that the errors are really small. We will conclude the maximal error in the 
   estimation of the asymptotic value of $p_{max}$ quite possibly is 0.0002 and therefore our final estimate
   is $p_{max} = 0.6033 \pm 0.0002$ which is distinctly different from the actual value of $p_c \approx 0.592746$.

\begin{figure}[t]
\includegraphics[width=6.0cm]{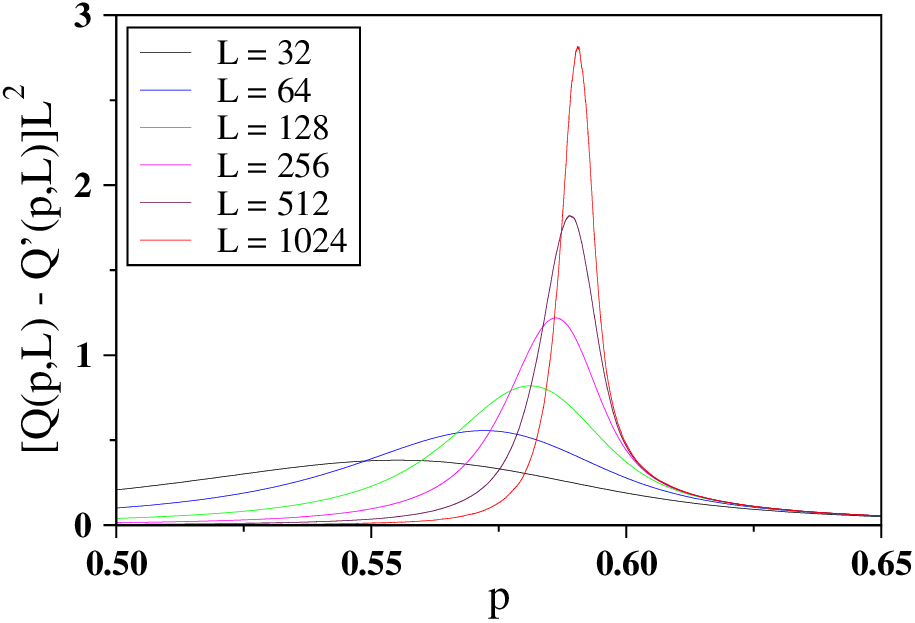}
\caption{
        The difference $\Delta Q = [Q(p,L) - Q'(p,L)]L^2$ have been plotted for six different sizes of the
        square lattice against the site occupation probability $p$. 
}
\label{FIG09}
\end{figure}

      We have also tried a log correction to the finite-size correction as follows:
\begin{equation}
\langle p_{max}(L) \rangle = p_{max} - AL^{-\alpha}(1-B(\ln L/L)).
\label {EQN02}
\end{equation}
   From our best fit we obtained $p_{max}$ = 0.60329, $A$ = 1.4323, $\alpha$ = 1.047 and $B$ = 0.2608
   which shows that the log correction has very little effect on $p_{max}$ (figure not shown).

      The collapse of the peak positions $Q_{max}(p_{max},L)$ of the curves in Fig. \ref {FIG05}
   on one another implies the maximal cluster size at this point $\langle s_{max}(L) \rangle \sim L^{\eta}$
   with $\eta$ = 2. This has been directly verified by plotting (figure not shown) $\langle s_{max}(L) \rangle$ against $L$
   using the double logarithmic scales for the square and simple cubic lattices. The values of the exponent 
   $\eta$ have been estimated from the slopes of the curves. For the square lattice $\eta_{sq} = 2.0057$ 
   has been obtained and for the simple cubic lattice $\eta_{sc} = 3.0075$ is found. This implies that 
   since $\langle p_{max} \rangle$ values are slightly larger than the percolation thresholds, the largest
   clusters turn out to be compact, and not fractals like the percolating clusters at the percolation thresholds.
   Consequently, their dimensions are equal to their embedding space dimensions. This analysis gives
   another support to our claim that $p_c$ and $p_{max}$ are indeed distinctly different.

      An alternate definition of $Q$-factor is as follows:
\begin{equation}
Q'(p,L) = \langle (s_{max}(p,L) / s_{av}(p,L)) \rangle / L^d.
\label {EQN03}
\end{equation}
   Here, for each value of $p$ of every run, one first calculates the quotient of the largest cluster size $s_{max}(p,L)$
   and the average cluster size $s_{av}(p,L)$ and then takes an average of this quotient over a large number of
   independent runs.

      We have calculated both the $Q(p,L)$ and $Q'(p,L)$-factors for the same set of runs. When we 
   plot these two $Q$-factors against $p$ on the same graph, it appears with the naked eye that one 
   curve completely overlaps the other as if the two factors are equal. Actually this is not the 
   case which becomes apparent when we plotted the difference $\Delta Q = [Q(p,L) - Q'(p,L)]L^2$ against 
   $p$ in Fig. \ref {FIG09} for six different sizes of the square lattice. It is observed that though 
   the maximal value of the difference is very small, there is a nice peak for $\Delta Q$ occurring 
   at $\langle p_{max}(L) \rangle$. The number of independent runs varied from $10^8$ up to $L$ = 64 
   to 320000 for $L$ = 1024. The locations of the maxima i.e., $p_{max}$ and $p'_{max}$ for $Q(p,L)$ 
   and $Q'(p,L)$ respectively are almost always the same, if not, they differ by an amount $\sim 1/L^2$.
   They are extrapolated as $\langle p_{max}(L) \rangle = 0.603312 - Const.L^{-2.1429}$.

\begin{figure}[t]
\includegraphics[width=6.0cm]{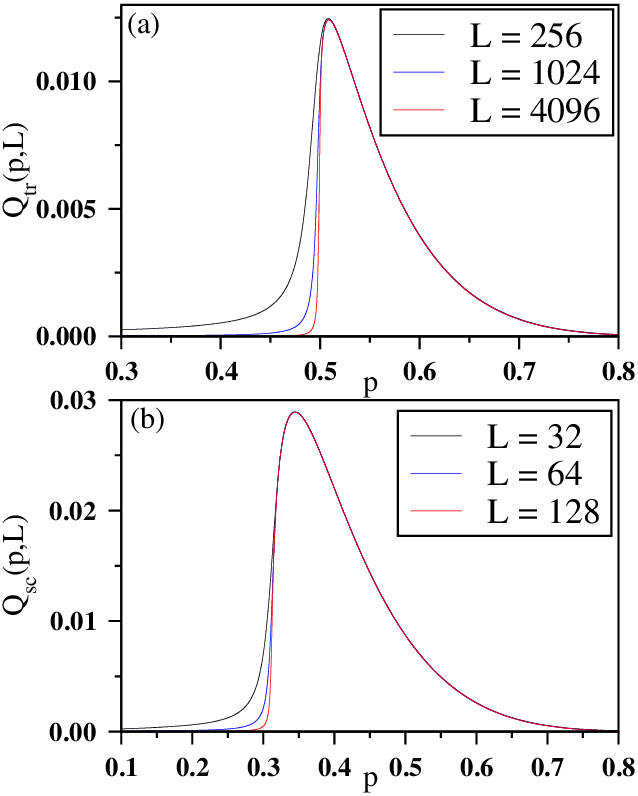}
\caption{Plots of the $Q$-factors against the site occupation probability $p$ for two different lattices:
   (a) $Q_{tr}(p,L)$ for the triangular lattice and
   (b) $Q_{sc}(p,L)$ for the simple cubic lattice.
}
\label{FIG10}
\end{figure}

\subsection {Triangular lattice}

      A parallel set of calculations have been done on the triangular lattice. A plot of $Q_{tr}(p,L)$ 
   against $p$ for three different system sizes have been shown in Fig. \ref {FIG10}(a). The positions of the 
   maxima are very close to the triangular lattice percolation threshold $p_c$ = 1/2 but slightly larger 
   than 1/2. For each run we have estimated the maximum value of $Q_{max}(p_{max},L)$, the corresponding 
   $p_{max}$ values and then averaged over all runs. The average $\langle p_{max}(L) \rangle$ values of 
   six different system sizes from $L$ = 128, ..., 4096 have been extrapolated to their asymptotic limit: 
   $\langle p_{max}(L) \rangle = p_{max} - const.L^{-1.053}$ with $p_{max}$ = 0.5088 
   which is approximately 0.9 percent different from the percolation threshold $p_c = 1/2$, see Fig. 
   \ref {FIG11}(a).

\subsection {Simple Cubic lattice}

\begin{figure}[t]
\includegraphics[width=6.0cm]{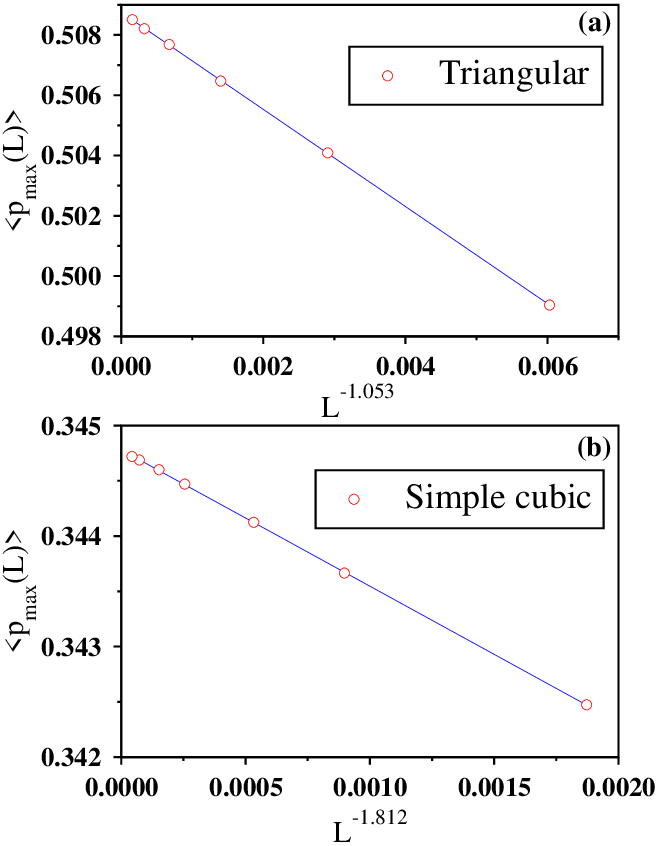}
\caption{
   Extrapolation of $\langle p_{max}(L) \rangle$ values to their asymptotic limit of $L \to \infty$
   gives the estimates of $p_{max}$: (a) 0.5088 for the triangular lattice, and (b) 0.3448 
   for the simple cubic lattice.
}
\label{FIG11}
\end{figure}

      For the simple cubic lattice we could study only small lattice sizes up to $L$ =256 and plot them in
   Fig. \ref {FIG10}(b). The only difference for the simple cubic lattice is the quotient of the maximal size 
   and average size is scaled by $L^3$, that is the total number of lattice sites in the system.
\begin{equation}
Q_{sc}(p,L) = (\langle s_{max}(p,L) \rangle / \langle s(p,L) \rangle) / L^3.
\label {EQN04}
\end{equation}
   The data for the positions of the maximum of $Q_{sc}(p,L)$ for the lattice sizes $L$ = 32 to 256 have 
   been used to extrapolate $\langle p_{max}(L) \rangle = p_{max} - const.L^{-1.812}$ 
   with $p_{max}$ = 0.3448 (Fig. \ref {FIG11}(b)). 

      Therefore, in each of the three lattices, namely, square, triangular and simple cubic
   we see that the precise values of the probabilities $p_{max}$ are about 1\% larger
   than their corresponding percolation thresholds $p_c$. Clear power laws for the finite-size 
   extrapolations in Fig. \ref {FIG06} and in Fig. \ref {FIG11} in all three cases indicate that indeed these
   threshold values $p_{max}$ are distinctly different from their $p_c$ values. These
   extrapolations are characterized by the exponents whose values are very are close, namely, 
   1.041 for the square lattice and 1.053 for the triangular lattice and widely different 
   1.812 for the simple cubic lattice which may be the indication of universality of the
   finite-size correction exponent. It may be that more extensive study in future with much larger 
   systems would yield values 1 and 2 for these exponents in two and three dimensions, a 
   possibility which we cannot rule out at this moment. Our conclusion that $p_{max}$ 
   are different, has also been supported by the independent measurements of the average mass of the 
   largest clusters at $p_{max}(L)$ which yield that indeed these clusters are of compact 
   structures instead of being fractals at their percolation thresholds. Here we like 
   to recall another problem of percolation connectivity between two points at distance of 
   separation of the order of the system size \cite {Manna-Ziff}. There also enhanced thresholds 
   for the percolation connectivities of the modified structure have been observed.

\begin{figure}[t]
\includegraphics[width=6.0cm]{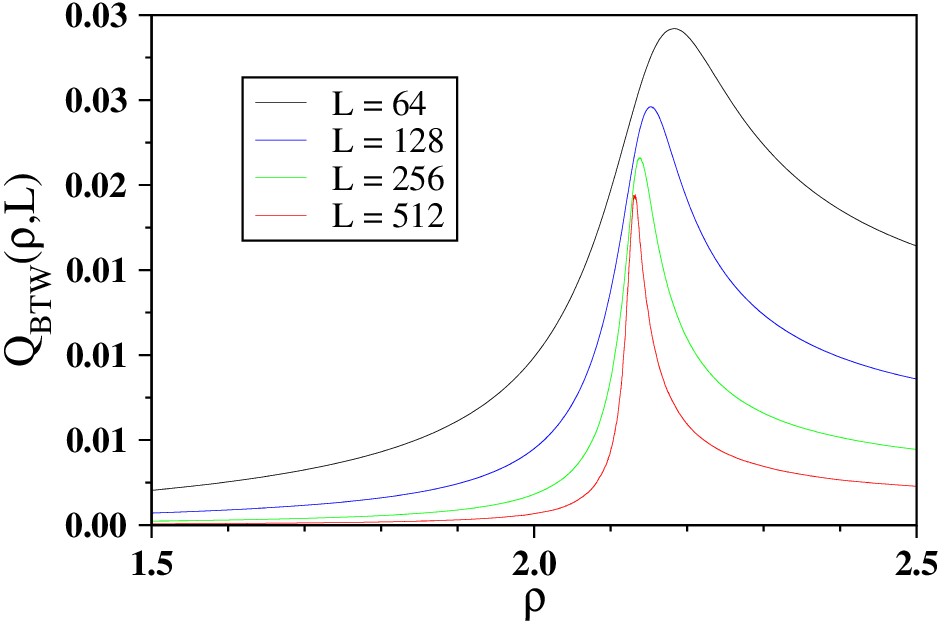}
\caption{
      For the BTW sandpile the values of $Q_{BTW}(\rho,L)$ have been plotted against the average
   number of sand particles $\rho$ dropped per site of a square lattice. 
}
\label{FIG12}
\end{figure}

\section {BTW sandpile}

      The Bak, Tang, and Wiesenfeld (BTW) sandpile \cite {btw} has been studied on the square 
   lattice of size $L \times L$ with open boundary condition. The dynamical evolution of the sandpile starts 
   from a completely empty lattice. Sand particles are dropped one by one at randomly selected 
   lattice sites. The system is allowed to relax through the deterministic BTW sandpile dynamics 
   \cite {btw}. The avalanche created by dropping one particle has size $s$, measured by the 
   total number of sand column topplings in the avalanche. At any arbitrary intermediate stage 
   of the sandpile dynamics, let $\rho$ be the average number of sand particles dropped per 
   lattice site. We refer it as the `drop density' which is a measure of the net inward current of sand 
   mass. This implies that an average number of $\rho L^2$ particles have been dropped on 
   to the system, many of which have left the system by jumping outside through the boundary. 
   Therefore, the drop density $\rho$ of particle addition is the control variable in this problem. 

      We have kept track of the maximal size $s_{max}(\rho,L)$ of all the avalanches created till 
   $\rho L^2$ particles have been dropped. At the same time we have also calculated the cumulative 
   average size $s_{av}(\rho,L)$ of all the avalanches of sizes larger than zero, including the 
   largest avalanche. Each run consists of a sequence of particle drops till the system moves well
   inside the stationary regime. We have checked that running the simulation till the drop density reaches a value of
   $\rho$ = 2.5 ensures arrival to the stationary state. Quantities which are averaged over many such independent runs are denoted by the 
   angular brackets $\langle ... \rangle$. Finally, we have defined a $Q$-factor which is the quotient 
   of the largest avalanche size and the average avalanche size of the sandpile:
\begin {equation}
Q_{BTW}(\rho,L) = (\langle s_{max}(\rho,L) \rangle / \langle s_{av}(\rho,L) \rangle) / L^2
\label {EQN05}
\end {equation}
   and plot this quantity against the drop density $\rho$ in Fig. \ref {FIG12}. There is a nice peak of 
   of value $Q_{max}(L)$ at the position $\rho_{max}(L)$ which we measure for four different system sizes.
   The rise and fall of $Q_{BTW}(\rho,L)$ values on the two sides of $\rho_{max}(L)$ are found to asymmetric.
   Therefore, the variation of $Q(\rho,L)$ against $\rho$ around the drop density $\rho_{max}(L)$ has a 
   $\lambda$-shape and the peak becomes sharper as the system size becomes larger. In the sub-critical 
   regime, sizes of all avalanches are small, so the value of $Q$ is small and $\sim 1$. The moment the 
   system moves into the stationary regime a very large avalanche abruptly appears which is quite generic 
   in all sandpile models. This makes the value of $s_{max}$ in the numerator quite large, but in comparison 
   the value of $s_{av}$ in the denominator increases only a little since all the avalanches
   share this increase in the total sum of all the avalanches. This results a rapid increase of $Q$. 
   Beyond the drop density $\rho_{max}(L)$ the system moves into the stationary state where the $s_{max}$
   increases very slowly, but $s_{av}$ increases very fast to reach a steady value. This ensures that after the peak $Q(\rho,L)$
   takes a stationary value as both the numerator and denominator assume steady values. This explains the
   $\lambda$-shape of the peak.

\begin{figure}[t]
\includegraphics[width=6.0cm]{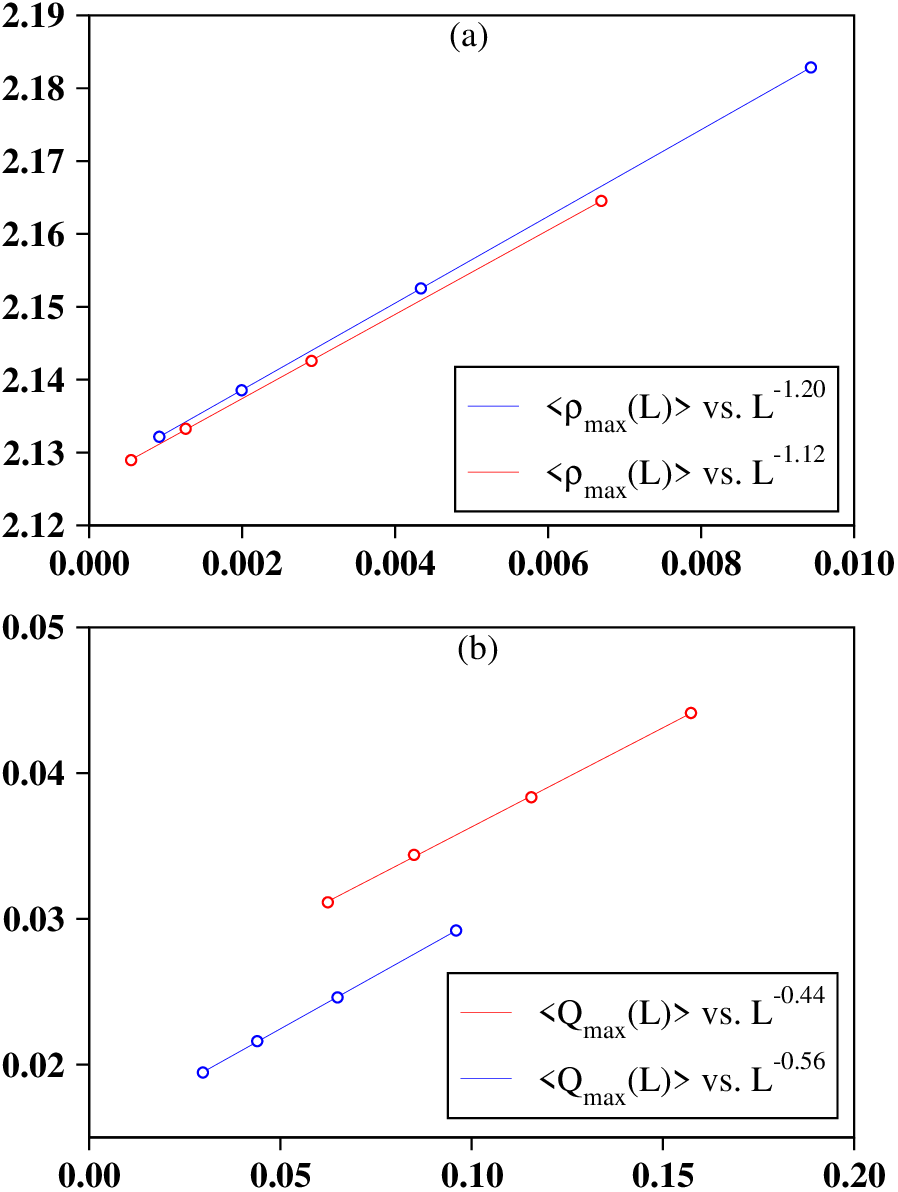}
\caption{
      For the BTW sandpile 
   (a) the drop densities $\langle \rho_{max}(L) \rangle$ for the maximal $Q$-factors and
   (b) the average values of the maximal $Q$-factors $\langle Q_{max}(L) \rangle$ have been plotted against 
   different negative powers of $L$. The two colors, red and blue, represent two different ways of calculations. 
   The extrapolated values in the asymptotic limit of $L \to \infty$ are consistent with each other.
}
\label{FIG13}
\end{figure}

      For a single sequence of sand grain additions on a system of size $L$, let $\rho_{max}(L)$ 
   be the precise value of the average number of particles dropped per site of the lattice 
   corresponding to the maximum value $Q_{max}(\rho,L)$ of the $Q$-factor of Eqn. \ref {EQN05}. 
   We have calculated $\langle \rho_{max}(L) \rangle$ using two different methods and plotted
   them against two different negative powers of $L$ in Fig. \ref {FIG13}(a).
   (i) Red: For each run we have we have picked up the maximum value $Q_{max}(\rho,L)$ of $Q$ 
   and its corresponding drop density $\rho_{max}$. These two quantities are then averaged over 
   a large number of independent runs. It has been observed that both $\langle \rho_{max}(L) 
   \rangle$ and $\langle Q_{max}(\rho,L) \rangle$ depend on the system size $L$ (Fig. \ref {FIG13}(b)). The best values 
   of the exponents for extrapolation of these two quantities are selected using the least square 
   fit method and the straight lines are then extrapolated to the $L \to \infty$ limit. We found 
   $\langle \rho_{max}(L) \rangle = 2.127 + 5.95L^{-1.12}$ and $\langle Q_{max}(L) = 0.023 + 0.136 L^{-0.44}$.
   (ii) Blue: Using a large number of runs and we have calculated for each value of $\rho$ the value of 
   $\langle s_{max}(\rho,L) \rangle$ and $\langle s_{av}(\rho,L) \rangle$ and then calculated $Q(\rho,L)$
   using Eqn. \ref {EQN05}. The maximum value $Q_{max}(\rho,L)$ have been determined and its location 
   $\rho_{max}(L)$ have been estimated. They are again best fitted by the least square method and 
   then extrapolated. We found $\langle \rho_{max}(L) \rangle = 2.126 + 5.78L^{-1.20}$ and 
   $\langle Q_{max}(L) \rangle = 0.015 + 0.147 L^{-0.56}$.

\section {Summary}

      How far the topper was ahead of a typical student in your class? One way to answer this question may
   be possible by looking at the total marks obtained in the final examination. Similarly how the richest in the society is
   ahead of the average members can be estimated by looking at their wealth. Thirdly, how the most famous research paper of a
   reputed scientist enjoys the maximum credit compared to the average credit of all his papers can be gauged by looking at his
   updated list of citation indices. Quite possibly one can cite more examples where the credit of the topper is
   compared with the average credit of the of a typical individual. 

      All these examples are dynamic in nature, e.g., the identification of the topper and the marks secured by him changes 
   from one exam to the other. Identity of the richest may also change from a year to the next, and so also the citations 
   received by the best paper of the scientist. Therefore, we thought it to be the best to consider a so-called `competition' 
   between the topper and the average and quantify this by defining the quotient of their credits as the $Q$-factor. The 
   natural question that comes to the mind, is why this study is important at all? The reason is this factor is a quantitative 
   measure of the fluctuations of the marks obtained by the students, wealth possessed by different members of the society,
   or the quality of the papers written by the scientist.

      We would like to recall that the citation statistics of majority scientists indicated growth of fluctuations in citations    
   with time. For very ``successful'' scientists, the statistical measures seem to indicate that these fluctuations reach an 
   universal SOC level. It was also observed that for a successful scientist the ratio of the citation number of the highest 
   cited paper to the average citation of all his papers often takes a value beyond a threshold (peak) value. In comparison,
   the value of the same ratio for not so reputed scientists do not reach that desired level. 

      This observation gave us the clue that the behavior of quotient of the largest to the average credits may be interesting to
   study in other physical systems as well. Therefore, in this paper we have decided to apply this idea to systems well known in 
   statistical physics. One example is the problem of percolation from the equilibrium systems and the other one is the sandpile
   model of self-organized criticality from the non-equilibrium systems. Both the systems evolve under suitably defined dynamical 
   rules. One defines the `connected clusters' in the percolation process and the `avalanche clusters' in the sandpile model 
   analogous to the group of members in the society. Though these systems in their early stages are un-correlated, under the 
   process of evolution they gradually become correlated. The signature of the correlation is traced in the rapid growth of the 
   largest cluster in the percolation process and the largest avalanche in the sandpile model. The credits possessed by these 
   members are estimated by the cluster sizes and the avalanche sizes. 

      In both the examples, the system passes through a transition point. On increasing the site occupation probability the
   percolating system makes a transition from the sub-critical to super-critical phase through the critical point. At this 
   point the size of the largest cluster grows at the fastest rate compared to the average size of all cluster. However,
   immediately after the percolation transition the rate of growth of the largest cluster slows down. As a consequence the 
   $Q$-factor exhibits a peak at a specific value of the site occupation probability $p_{max}$ which is about $\sim 1\%$ 
   larger than the percolation threshold $p_c$ of all lattices. We have argued in the text that these two numbers $p_{max}$ 
   and $p_c$ cannot be the same, only because of the fact that the percolation transition is a continuous transition. A very 
   similar scenario arises for the sandpile model where the current size of the largest avalanche undergoes a large jump in its 
   size when the system moves into the self-organized stationary state.

\section*{Acknowledgement}

      We are very much thankful to Amnon Aharony and Robert Ziff for their insightful comments and suggestions. BKC
   is grateful to the Indian National Science Academy for their Senior Scientist Research Grant.

\end {document}